# Study of the composition dependence of the ionic conductivity of LiH$_x$D$_{1-x}$ alloys

Vassiliki Katsika-Tsigourakou

*Section of Solid State Physics, Department of Physics, National and Kapodistrian University of Athens, Panepistimiopolis, 157 84 Zografos, Greece*

**Abstract**

Recent observations have shown that at a certain composition of solid solutions AgCl$_x$Br$_{1-x}$, AgCl-CdCl$_2$, AgBr-CdBr$_2$, of ionic crystals, the electrical conductivity exhibits a value appreciably larger than that of the end constituents. Here, we investigate the electrical conductivity σ of LiH$_x$D$_{1-x}$ solid solutions –which are of prominent importance in fuel hydrogen storage applications- and find that in the whole composition range no maximum is likely to occur in the σ versus *x* dependence.

---

e-mail: vkatsik@phys.uoa.gr

# 1. Introduction

In general, the study of the hydrides LiD and LiH has attracted a strong interest in view of their tentative important technological applications such as in fuel hydrogen storage. Much work has been made on the high pressure properties of the LiH isotopes, for two reasons: First, the comparison between theory and experiment in this case constitutes an important test of the techniques employed in the electronic structure calculations [1]. Second, the data of these isotopes are of chief importance in nuclear fusion research. Synchrotron X-ray diffraction measurements have been made [2] on solid $^7$LiH up to 36 GPa and solid $^7$LiD up to 94 GPa by using helium as a hydrostatic medium and reported that the rock-salt structure is stable up to the maximum pressure and that at ambient conditions the isothermal bulk modulus is $B_I$=32.2 GPa and its pressure derivative $dB_I/dP$=3.53. These measurements also found an isotopic shift on the equation of state the magnitude of which falls in between various approximations of the contribution of the zero-point lattice vibrations. Note also that, in the conductivity ($\sigma$) measurements [3] in $^7$LiH and $^8$LiD an isotopic shift has been also observed.

Recent conductivity measurements [4] as a function of temperature and composition have been reported on solidified mixtures [5-7] of AgCl-CdCl$_2$ and AgBr-CdBr$_2$. These solid solutions exhibited an important increase in conductivity with the concentration of Cd$^{2+}$, until around 20 mol% CdX$_2$ (X=Cl, Br). At the maximum, the value of electrical conductivity in the AgCl+ CdCl$_2$ solid solution is about 40 times higher than in pure AgCl and the value of electrical conductivity in AgBr+CdBr$_2$ solid solution is about 3 times higher than in pure AgBr. A quantitative explanation of this phenomenon, i.e., the appearance of the maximum conductivity of the solid solution at a certain concentration of Cd$^{2+}$, has been recently treated by Skordas [8] by employing an established thermodynamical model –termed cBΩ model [9-11] (see below). Note that the same model has been employed [12] for the explanation of the emission of electric signals before fracture –at ionic solids upon gradually increasing stress- which provides the basis for the observation of precursory signals before earthquakes [13-16].

It is the scope of this paper to use this thermodynamical model to investigate the electrical conductivity of $LiH_xD_{1-x}$ versus x. Hence, it is our aim to identify whether the aforementioned phenomenon (i.e., a maximum conductivity of a certain value of x) may appear in the system $LiH_xD_{1-x}$.

## 2. Estimation of the composition of the maximum conductivity.

Following Skordas [8] and considering aspects on defects [17,18], if $V_I$ and $V_{II}$ denote the molar volumes of the two pure constituents (*I*) and (*II*), the "molar" volume *V* of the solid solution can be written as [16]:

$$V = V_I(1-x) + V_{II}x \qquad (1)$$

where *x* stands for the molar concentration of the crystal *II* in the mixed system. Differentiating Eq.(1) with respect to pressure, we find [16]:

$$\frac{B}{B_I} = \frac{1 + x\dfrac{Nv^d}{V_I}}{1 + x\dfrac{\kappa^d}{\kappa_I}\dfrac{Nv^d}{V_I}} \qquad (2)$$

where *B* and $B_I$ are the bulk modulus of the solid solution and the pure crystal *I*, respectively, and $\kappa_I = 1/B_I$ where $\kappa_I$ is the compressibility of the pure crystal *I*. The quantity $v^d$ stands for a defect volume which may be identified as follows: It represents the difference of the volume of a crystal of $N$ molecules of type $I$ and the same crystal in which one of its "molecules" has been exchanged by a molecule of type $II$. Assuming that $V_{II} > V_I$ we get [16]:

$$Nv^d = V_{II} - V_I \tag{3}$$

As for, the quantity $\kappa^d$ in Eq.(2), it represents the compressibility $\kappa^d$ of the defect volume $v^d$:

$$\kappa^d = -\frac{1}{v^d}\frac{dv^d}{dP}\bigg|_T \tag{4}$$

This differs in general from the compressibility $\kappa_I$ of the component $I$, i.e.,

$$\frac{\kappa^d}{\kappa_I}(\equiv \mu) \neq 1 \tag{5}$$

A combination Eqs.(2), (3) and (5) leads to

$$\frac{B}{B_I} = \frac{1+x\lambda}{1+x\mu\lambda} \tag{6}$$

where the symbol $\lambda$ stands for:

$$\lambda \equiv \frac{V_{II}}{V_I} - 1 > 0 \tag{7}$$

Combining Eqs.(1) and (6) we find:

$$\frac{BV}{B_I V_I} = \frac{(1+x\lambda)^2}{1+x\mu\lambda} \qquad \text{or}$$

$$\frac{B\Omega}{B_I\Omega_I} = \frac{(1+x\lambda)^2}{1+x\mu\lambda} \tag{8}$$

where the symbols $\Omega$ and $\Omega_I$ denote the mean volume per atom of the mixed system and the "pure" crystal $I$, respectively.

According to the cBΩ model the defect Gibbs activation energy $g^{act}$ is given by [9-11]:

$$g^{act} = c^{act}B\Omega \tag{9}$$

where $c^{act}$ is independent of temperature and pressure. Thus, we can write:

$$\frac{g^{act,x}}{g^{act,I}} = \frac{c^{act,x}}{c^{act,I}}\frac{B\Omega}{B_I\Omega_I} \tag{10}$$

where $g^{act,x}$ and $g^{act,I}$ stand for the activation Gibbs energies for the mixed crystal and the pure crystal ($I$), respectively.

Assuming that $c^{act,x}$ varies only slightly versus the composition, Eq.(10) leads to

$$\frac{g^{act,x}}{g^{act,I}} = \frac{(1+x\lambda)^2}{1+x\mu\lambda} \tag{11}$$

We now consider that the conductivity σ, for a single conduction process, is given [16] by:

$$\sigma \propto \exp\left(-\frac{g^{act}}{kT}\right) \tag{12}$$

Assuming that the preexponential factor in Eq.(12) does not change markedly with composition, we observe that the variation of the conductivity with the composition is governed by the function $(1+x\lambda)^2/(1+x\mu\lambda)$ appearing in Eq.(11). This function reaches a minimum value when the molar concentration $x$ of the constituent $II$ takes the value

$$x_m = \frac{\mu-2}{\lambda\mu} \qquad (13)$$

The quantity $\lambda$ is obtained from Eq.(7) because the molar volumes of the end members are always known. As for the quantity $\mu$ it can be estimated by the aforementioned thermodynamical model by the following procedure: Substituting Eq.(9) into the well known [16] relation $v^d = \left.\dfrac{dg^{act}}{dP}\right|_T$, we find $v^d = c^{act}\left[\left.\dfrac{dB}{dP}\right|_T - 1\right]\Omega$, which when inserted into Eq.(4), gives:

$$(\mu \equiv)\frac{\kappa^d}{\kappa_I} = 1 - \frac{B_I \left.\dfrac{d^2 B_I}{dP^2}\right|_T}{\left.\dfrac{dB_I}{dP}\right|_T - 1} \qquad (14)$$

The quantities $dB_I/dP$ and $d^2B_I/dP^2$, when they are not experimentally accessible, can be estimated from the modified Born model according to [16]:

$$dB_I/dP = (n^B + 7)/3 \qquad (15)$$

and

$$B_I(d^2B_I/dP^2) = -(4/9)(n^B + 3) \qquad (16)$$

where $n^B$ is the well known [16] Born exponent.

## 3. Application of the aforementioned methodology to $LiH_xD_{1-x}$.

Using the lattice parameters 4.084 Å for LiH and 4.060 for LiD (see Ref. 3 and references therein), Eq.(7) leads to

$$\lambda = 0.018 \qquad (17)$$

By applying Eq.(15) and using the aforementioned value $dB_I/dP = 3.53$ we find $n^B = 3.6$. This Born exponent, when inserted into Eq.(16) –upon using also the value $B_I = 32.2$ GPa- results in $(d^2 B_I/dP^2)_T = -0.09$ GPa$^{-1}$. By substituting this value into Eq.(14) we get

$$\mu = 2.17 \qquad (18)$$

We now turn to Eq.(13) which reveals that, since $0 \leq x_m \leq 1$, the observation of a maximum value in the curve $\sigma$ versus $x$ demands:

$$0 \leq \frac{\mu - 2}{\lambda \mu} \leq 1 \qquad (19)$$

An inspection of the inequality (19) –in view of the $\lambda$, $\mu$ values determined above in the relations (17) and (18)- leads to the conclusion that this inequality is violated.

## 4. Conclusion

Here, we investigating whether the electrical conductivity $\sigma$ of the solid solutions $LiH_xD_{1-x}$ exhibits a maximum value at a certain composition, which has been experimentally observed in some other ionic mixed systems. We conclude that this is not the case for $LiH_xD_{1-x}$, which may be of high importance for future fuel hydrogen storage research.